\documentclass[aip,amsmath,amssymb,reprint]{revtex4-2}
\usepackage{graphicx}
\usepackage{tabularx}
\usepackage{booktabs}
\usepackage[utf8]{inputenc}
\usepackage[T1]{fontenc}
\usepackage[pdfstartview=FitH, pdfpagemode=None, pdfpagelayout=OneColumn, colorlinks=true, linkcolor=blue, citecolor = blue]{hyperref}

\makeatletter
\def\@email#1#2{%
 \endgroup
 \patchcmd{\titleblock@produce}
  {\frontmatter@RRAPformat}
  {\frontmatter@RRAPformat{\produce@RRAP{*#1\href{mailto:#2}{#2}}}\frontmatter@RRAPformat}
  {}{}
}%
\makeatother

\begin{document}
\title{Features of long particle beam self-modulated in plasma}
\author{V.M. Yarygova}
\author{K.V. Lotov}%
\affiliation{ 
Budker Institute of Nuclear Physics, Novosibirsk, 630090, Russia
}%

\date{\today}

\begin{abstract}
Under certain conditions, a long particle beam self-modulates in a dense plasma, that is, it breaks down into a train of short, stable microbunches under the influence of its own wakefield.
During this process, the beam also changes its radial profile: initially having a Gaussian shape, it evolves to a highly peaked equilibrium state with a density singularity on the axis.
This change makes individual beam slices several times more efficient at exciting the wakefield.
We have developed an analytical iterative model describing the transverse equilibrium state of the microbunches, which is applicable to most beam cross-sections.
The model is based either on the conservation of the transverse adiabatic invariant or on empirically established relationships in cases where adiabaticity is violated. 
It predicts the radial profiles of beam density and wakefield potential, as well as the particle distribution in the transverse phase space.
The model is benchmarked against numerical simulations and demonstrates a high degree of  accuracy.
In addition to the full model, we present simplified engineering formulas based on elementary functions that bypass iterative procedures.

\end{abstract}

\maketitle

\section{Introduction}

Plasmas can withstand strong electric fields, orders of magnitude stronger than those that can be achieved in solid-state structures.\cite{PoP27-070602,RMP81-1229}
These fields can accelerate, 
decelerate,\cite{PRST-AB13-101303, PoP22-083106, PPCF61-124002} 
focus,\cite{PAcc20-171, PoP2-2555, PRAB19-071301, PoP17-073105, PRL115-184802, Nat.530-190, PRL121-174801, PRL123-054801, PRAppl22-034022} 
steer\cite{PRST-AB4-091301, PRL132-215001} 
charged particle beams, or change their shape.
In particular, the plasma fields can slice long beams into short bunches.\cite{EPAC98-806, PRL104-255003}
The bunched beams can be used to generate terahertz radiation\cite{PRST-AB15-111301, PoP25-043111, AIPAdv9-025025} 
or, more commonly, to effectively excite waves for wakefield acceleration.\cite{PRL54-693}
Plasma wakefield acceleration with bunch trains is being experimentally investigated in several laboratories around the world,\cite{PPCF61-045012, PRL112-045001, PoP29-100701, PRAB24-051302} 
including CERN, where the AWAKE experiment is studying wave excitation by a proton beam.\cite{Nat.561-363, Symmetry14-1680}

To predict and analyze the results of experiments, it is necessary to know the shape and momentum distribution of the bunches that excite the wave.
The beam inside the plasma is subject to strong focusing, so the particles are pulled towards the axis and their density distribution changes.
Calculating the plasma response on the basis of the initial beam density profile (e.g., Gaussian\cite{PoP12-063101}) may lead to an incorrect evaluation of the plasma fields.

There are several models available of beam radial equilibrium in plasma based on different assumptions about the particle distribution in phase or coordinate space: Gaussian distribution in momentum\cite{PFB2-1376, PRE49-4407} 
or coordinates,\cite{PoP2-1326, PRAB20-111301} 
flat-top beam profile in the radial direction,\cite{PoP23-053109, LPB34-519}
or a special law governing the change in the amplitudes of radial oscillations of particles.\cite{PoP24-023119}
Only the model of Ref.\,\onlinecite{PoP24-023119} demonstrates good quantitative agreement with high-resolution simulations, but its applicability is limited to single short bunches.

In this paper, we develop analytical models describing the equilibrium state reached by particle beams in a dense plasma. 
The models allows calculating the radial distributions of the beam density and the wakefield potential, as well as the beam density in the transverse phase space, without simulating the transition to this equilibrium state with any numerical code. 
After describing the research method in Sec.~\ref{sec:methods}, we first examine, in Sec.~\ref{sec:phenom}, the main features of beam equilibrium identified in numerical simulations. 
Next, in Sec.~\ref{sec:analyt}, we develop analytical models applicable to different parts of the beam and check their accuracy by comparing them with the results of numerical simulations.
In Sec.~\ref{sec:approx}, we present less rigorous but easy-to-use engineering formulas for describing the equilibrium state of a beam, which are nevertheless quite accurate.
In Sec.~\ref{sec:discuss}, we discuss the applicability of the results obtained and the sources of inaccuracies.
In Sec.~\ref{sec:summ}, we summarize the main findings.

\section{Methods}
\label{sec:methods}

We use the following research method. 
First, we simulate a test problem that helps us understand the main features of beam self-modulation in plasma. 
Based on these, we then formulate an analytical model, benchmark it against simulations, and, where possible, write down simplified, easy-to-use formulas for the key features of the equilibrium beam state.

The effects we study are axially symmetric and related to the beam, so we use the co-moving coordinate $\xi = z-ct$ along with the usual cylindrical coordinates $(r, \varphi, z)$, where $\vec{z}$ is the direction of beam propagation and $c$ is the speed of light.
The natural units for the problem are determined by the plasma density $n_0$: $\omega_p = \sqrt{4 \pi n_0 e^2/m}$ for frequencies, $k_{p0}^{-1} = c/\omega_p$ for distances, and $E_0 = m c \omega_p/e$ for fields, where $m$ is the electron mass and $e>0$ is the elementary charge.

\begin{table}[tb]
   \centering
   \caption{Parameters of the test problem}
   \begin{tabular}{lll}
       \toprule
       Parameter, notation && Value \\
       \midrule
        Beam density, $n_{b0}$         &&         $4 \times 10^{-3} n_0$\\
        Beam radius, $\sigma_r$       &&         $0.5 k_{p0}^{-1}$\\
        Beam relativistic factor, $\gamma_b$   &&         $1000$\\
        Beam angular spread, $\delta \alpha$       &&        $2 \times 10^{-6}$\\
        Beam length, $L_b$       &&        $160 k_{p0}^{-1}$\\
        Plasma length, $L_p$       &&        $10000 k_{p0}^{-1}$\\
        Position of plasma density step, $z_s$       &&        $360 k_{p0}^{-1}$\\
        Plasma density rise, $\delta n$       &&        $0.085 n_0$\\
       \bottomrule
   \end{tabular}
   \label{tab:params}
\end{table}

\begin{figure}[tb]
\includegraphics{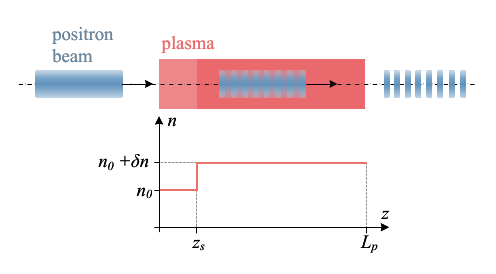}
\caption{ Outline of the test problem. The lower fragment schematically shows the longitudinal distribution of plasma density.}
\label{fig1-setup}
\end{figure}

In the test problem taken from Ref.~\onlinecite{PoP22-103110}, a positron beam of constant radius propagates in a radially uniform plasma with immobile ions (Table~\ref{tab:params} and Fig.\,\ref{fig1-setup}). 
Its initial density is
\begin{equation}\label{e1}
 n_b = \left\{ \begin{array}{ll}
 \displaystyle n_{b0} \, e^{-r^2/(2 \sigma_r^2)}, & -L_b < \xi < 0, \\
 0, & \xi \geq 0.
 \end{array}\right.
\end{equation}
The sharp leading edge of the beam creates a seed wakefield from which the self-modulation develops. 
Choosing positively charged beam particles makes the discussion of potential wells and hills more intuitive.
The beam density is low enough to neglect nonlinearities of the plasma response. 
The high relativistic factor allows us to clearly separate the timescales of plasma oscillations, transverse beam dynamics, and beam deceleration. 
The low angular spread eliminates effects related to beam emittance from consideration.\cite{PoP22-123107}

To prevent beam destruction by its own field at the final stage of self-modulation,\cite{PoP18-024501,PoP18-103101} the longitudinal plasma density profile must be non-uniform and contain a rise, called density step, at the linear stage of self-modulation.\cite{PoP22-103110}
There is currently no theoretical model that predicts the required parameters of the density step.
So they have to be found using multi-parameter numerical optimization, which can be quite computationally intensive.\cite{PPCF62-115025}
Here we use the density profile $n(z)$ from Ref.~\onlinecite{PoP22-103110}:
\begin{equation}\label{e2}
    n = \left\{ \begin{array}{ll}
 \displaystyle n_0, & z \leq z_s, \\
 n_0 + \delta n, & z > z_s.
 \end{array}\right.
\end{equation}

We simulate the beam dynamics with axisymmetric quasistatic code LCODE.\cite{PRST-AB6-061301,NIMA-829-350}
The radius of simulation domain is $8k_{p0}^{-1}$. 
The grid steps are $\Delta r = \Delta \xi = 0.01 k_{p0}^{-1}$, and the step for updating the beam state is $\Delta z = 10 k_{p0}^{-1}$.
There are $4 \times 10^7$ equal macro-particles in the beam. 
Plasma electrons are represented by 10 radius-weighted macro-particles per radial interval~$\Delta r$. 
Plasma ions are considered to be uniformly distributed positive charges.

It is convenient to characterize the combined action of electric ($\vec{E}$) and magnetic ($\vec{B}$) fields by the wakefield potential $\Phi$, the gradient of which determines the force exerted on an axially moving relativistic unit charge:
\begin{equation}\label{e3}
    \Phi = -\int E_z d \xi = -\int (E_r - B_\varphi) d r,
\end{equation}
and define $\Phi_0 = mc^2/e$.

If the plasma response to the beam is linear and the beam density is separable at some $z$,
\begin{equation}
    n_b (r, \xi) = n_{b0}\, n_\perp (r)\, n_\parallel (\xi),
\end{equation}
the wakefield potential is also separable:\cite{PAcc22-81}
\begin{gather}
    \label{e5}
    \Phi (r, \xi) = \Phi_0 \frac{n_{b0}}{n} Z(\xi) R(r), \\ 
     Z(\xi) = k_p \int_\xi^\infty d\xi' \, n_\parallel (\xi') \, \sin \bigl( k_p (\xi' - \xi)\bigr), \\
     \nonumber
     R(r) = - k_p^2 \int_0^r d r' r' I_0 (k_p r') K_0 (k_p r) \, n_\perp (r') \\
     \label{e7}
        - k_p^2 \int_r^\infty d r' r' I_0 (k_p r) K_0 (k_p r') \, n_\perp (r'),
\end{gather}
where $K_0$ and $I_0$ are the zero-order modified Bessel functions, and $k_p$ is the wavenumber corresponding to the local plasma density $n$.
For definiteness, we assume
\begin{equation}\label{e8}
    k_p^2 \int_0^\infty n_\perp (r) \, 2 \pi r \, dr = 1
\end{equation}
and its consequence (Appendix~\ref{app0})
\begin{equation}\label{e8a}
    k_p^2 \int_0^\infty R (r) \, 2 \pi r \, dr = -1,
\end{equation}
so the functions $n_\perp (r)$, $n_\parallel (\xi)$, $Z(\xi)$ and $R(r)$ are dimensionless.
For the initial beam profile \eqref{e1}, we have
\begin{equation}\label{e9}
    n_\perp (r) = \frac{e^{-r^2/(2 \sigma_r^2)}}{2 \pi k_p^2 \sigma_r^2}, \quad n_\parallel = 2 \pi k_p^2 \sigma_r^2.
\end{equation}

The bunches emerge in the plasma of initial density $n_0$, and their periodicity is determined by the plasma period there, that is why we use $k_{p0}^{-1}$ as the unit length when displaying longitudinal dependencies.

If the beam density is not strictly separable, we can still use formula \eqref{e7} to characterize the relationship between radial profiles of beam density and wakefield potential. 
For this case, we define
\begin{equation}\label{e10}
    n_\perp (r) = A \int n_b (r, \xi') \, d\xi',
\end{equation}
where the factor $A$ ensures that condition \eqref{e8} is satisfied, and the integration allows the beam density to be averaged over one or more intervals of $\xi$.

Contributions of individual beam slices to the wakefield formation are conveniently characterized by the dimensionless effective current\cite{PoP32-033102, PPCF64-075003}
\begin{equation}
    I_\text{eff} (\xi, z) = \frac{k_p^2}{n} \int_0^\infty K_0 (k_p r) n_b (r, \xi, z)\, r \, dr.
\end{equation}
It determines the on-axis wakefield potential in the linearly responding plasma, even if the beam density is not separable:
\begin{equation}
    \Phi (0, \xi, z) = - \Phi_0 k_p \int_\xi^\infty \, I_\text{eff} (\xi', z) \, \sin \bigl( k_p (\xi' - \xi)\bigr) \, d\xi'.
\end{equation}
If the beam density is separable at some $z$, then
\begin{equation}\label{e13}
    I_\text{eff} (\xi) = - \frac{n_{b0} n_\parallel (\xi)}{n} R(0).
\end{equation}

\begin{figure}[tb]
\includegraphics{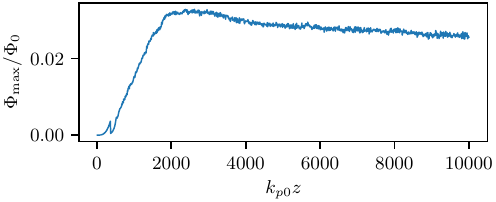}
\caption{Maximum wakefield potential $\Phi_\text{max}$ versus beam propagation distance $z$. }
\label{fig2-phi_max}
\end{figure}

\begin{figure*}[tb]
\includegraphics{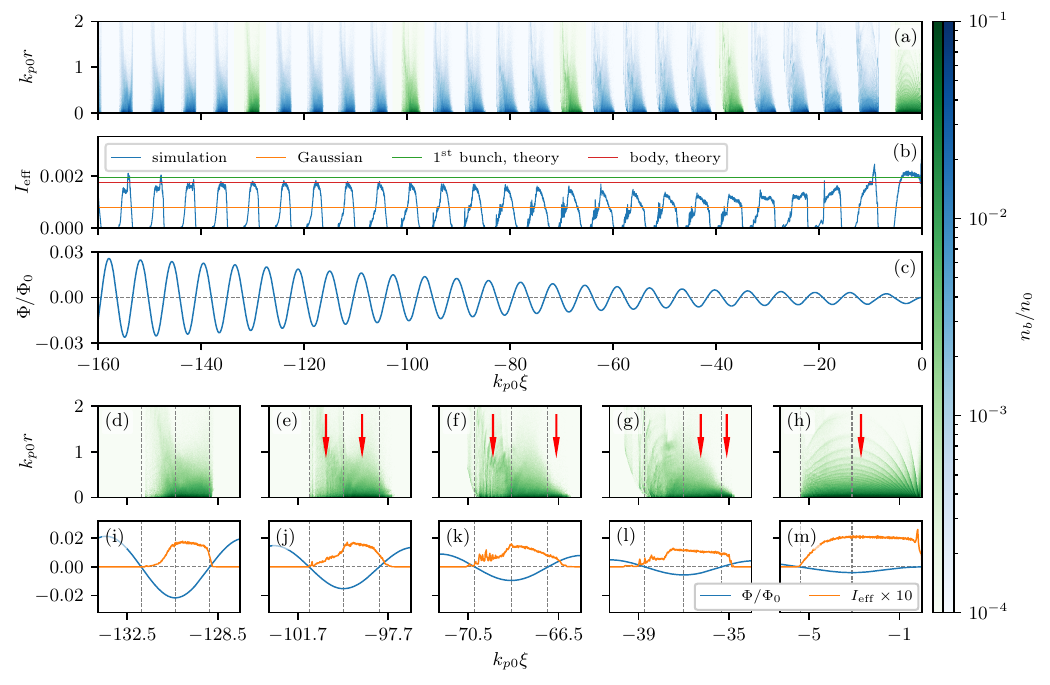}
\caption{ (a) Portrait of the self-modulated beam, (b) effective current $I_\text{eff}(\xi)$ of bunches and theoretical predictions for it described in the text, (c) on-axis wakefield potential $\Phi(\xi)$ of the excited wave, (d)-(h) portraits of individual bunches marked in fragment (a) with a green pallette, and (i)-(m) wakefield potential and effective current in their vicinity.
The vertical dashed lines in (d)-(m) help to distinguish between the regions of accelerating, decelerating, or defocusing wakefield. 
The red arrows in (e)-(h) indicate the beam cross-sections detailed in Fig.\,\ref{fig5-simphase}.
}
\label{fig3-general}
\end{figure*}

As the beam propagates in the plasma, the amplitude of the excited wakefield, characterized by the maximum wakefield potential $\Phi_\text{max}$ (Fig.\,\ref{fig2-phi_max}), first rapidly grows and then slowly decreases.
If the propagation distance $L_p$ is much longer than the scale of fast changes in the wakefield, we can consider the beam state at $z=L_p$ as the final result of self-modulation and study its features.
However, some beam slices, where the transverse fields are weak, approach the equilibrium state very slowly. 
Some particles that are defocused by the wave do not have enough time to move away from the beam.
Such particles lie in the phase space outside the separatrix, which is defined by the equation
\begin{equation}\label{e14}
    S(r,\xi) = \pm m c \sqrt{2 \gamma_b \bigl(\Phi_m(r,\xi) - \Phi(r, \xi)\bigr) / \Phi_0},
\end{equation}
where
\begin{equation}
    \Phi_m(r,\xi) = \max_{r' > r} \bigl(\Phi(r', \xi)\bigr)
\end{equation}
is the height of the potential hill that keeps particles from being defocused away.
A particle with coordinates $(r_p,\xi_p)$ and momentum $p_r$ is not confined radially if 
\begin{equation}
    |p_r| > |S (r_p, \xi_p)|.
\end{equation}
There are not many of them, so their presence does not affect the results if $k_{p0} L_p$ is sufficiently large.

\section{Phenomenology}
\label{sec:phenom}

The lengths of the bunches in the fully self-modulated beam varies, with earlier bunches being longer [Fig.\,\ref{fig3-general}(a)].
Despite the different focusing strengths at different beam cross-sections, the effective current does not change drastically along the bunch train [Fig.\,\ref{fig3-general}(b)] and along individual bunches [Fig.\,\ref{fig3-general}(i)-(m)].
This suggests that the rate of beam compression by the plasma wakefield also depends weakly on the focusing strength, at least in the central parts of the bunches [Fig.\,\ref{fig3-general}(d)-(h)].
The wakefield amplitude monotonically increases along the beam [from right to left in Fig.\,\ref{fig3-general}(c)]. 

In each bunch, we can distinguish three parts [Fig.\,\ref{fig3-general}(d)-(m)]: the body, being focused ($\Phi<0$) and decelerated ($\partial_\xi \Phi > 0$) by the wave; the head, located at $\Phi>0$; and the tail, being focused and accelerated.

The bodies have an almost constant effective current and transfer energy to the wave.

The heads survive because the radial dependence of the wakefield potential has an off-axis maximum that prevents near-axis particles from being defocused away.

The tails take energy from the wave.
They are not needed for efficient wave excitation, and it may be possible to reduce or avoid them with some sophisticated dependence $n(z)$.\cite{JPCS1067-042009} 
However, for plasma density profiles with a single density step, they are always present.
The tails can cause the effect of wakefield regeneration, when the wave restores its amplitude shortly after passage of the accelerated witness bunch.\cite{PRR7-L012055}

The first bunch has a different pattern, and later we will see that there are physical reasons for this.

\begin{figure}[tb]
\includegraphics{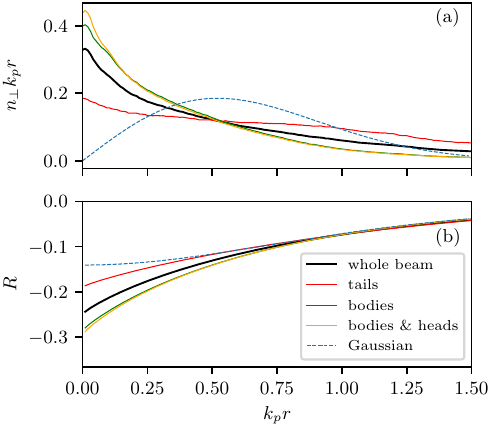}
\caption{ (a) $r$-weighted radial density profiles $n_\perp (r)$, averaged over different parts of the beam, and (b) the corresponding radial profiles $R(r)$ of the wakefield potential. 
The dashed lines correspond to the initial Gaussian profile \eqref{e1} of the beam.}
\label{fig4-shape}
\end{figure}

To get a general idea of the radial density distribution in the bunches, we average the beam density according to formula~\eqref{e10} over the bunch bodies, tails, heads, or their combinations [Fig.\,\ref{fig4-shape}(a)].
The density profiles $n_\perp (r)$ are so strongly peaked near the axis that even after multiplying by $r$ they reach their maximum value there.
This means that equilibrium bunches have a density singularity on the axis ($n_b \propto 1/r$).
Such a singularity makes them very different from Gaussian beams, which do not have this feature.

The singular behavior of the density leads to a funnel-shaped radial profile of the wakefield potential, in which the derivative $\partial_r R(r)$ tends to a nonzero constant as $r \to 0$ [Fig.\,\ref{fig4-shape}(b)].
A similar feature is observed for single bunches approaching radial equilibrium in high density plasma.\cite{PoP24-023119}

The bunch tails are, on average, wider and less peaked than the bodies  [Fig.\,\ref{fig4-shape}(a)], and their density patterns differ from bunch to bunch [Fig.\,\ref{fig3-general}(d)-(m)]. 
This is because the tails spend some time in the defocusing phase of the wave during the self-modulation process.\cite{PoP22-103110} 
In a hypothetical plasma with a density profile perfectly tailored for ideal self-modulation, there should be no bunch tails; they should be completely defocused. 
With the imperfect plasma density profile that we have, the tails are not completely defocused, but only partially.
Their particles are radially scattered during the defocusing time, resulting in various density patterns. 
For this reason, creating a general theory describing the bunch tails is hardly possible, so we focus on the bunch bodies.

\begin{figure}[tb]
\includegraphics{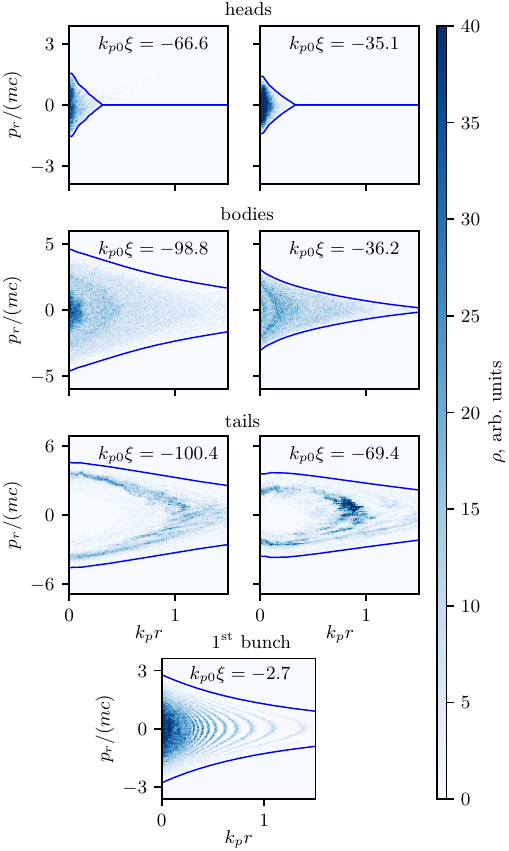}
\caption{ Beam density $\rho$ in the transversal phase space $(r, p_r)$ for several cross-sections indicated in Fig.\,\ref{fig3-general}.
For the heads, particles are displayed within intervals of length $0.2k_{p0}^{-1}$ near the specified values of $\xi$, and in all other cases, within intervals of length $0.8k_{p0}^{-1}$.}
\label{fig5-simphase}
\end{figure}

The number of particles in the bunch heads is small, despite their visually high density [Fig.\,\ref{fig3-general}(e)-(g)] and high effective current [Fig.\,\ref{fig3-general}(i)-(m)].
Contribution of these beam parts to the wave is also small, as we see from comparing green and yellow lines in Fig.\,\ref{fig4-shape}, so we neglect them.

Because of the funnel-shaped wakefield potential, the separatrixes defined by Eq.~\eqref{e14} in most cross-sections of the beam have a triangle-like shape (Fig.\,\ref{fig5-simphase}). 
All particles that survive during self-modulation are located inside the separatrixes.

We calculate the particle density in the phase space $\rho(r, p_r, \xi)$ (Fig.\,\ref{fig5-simphase}) as the number of beam particles in unit volume $dr\,d\xi\, dp_r$.
It is related to the beam density $n_b$ as follows:
\begin{equation}\label{e16}
    \int_{-\infty}^{\infty} \rho(r, p_r, \xi)\, dp_r = 2\pi r n_{b}(r, \xi).
\end{equation}
According to its definition, this function contains a spatial geometrical factor $2\pi r$, which suppresses it at small $r$.
Despite this, in the beam layers that transfer energy to the wave (bodies, heads, first bunch), the phase density $\rho$ reaches its maximum at $r = 0$.
The particle distributions in the bunch tails have quite a specific structure, in which most particles oscillate with large amplitudes and are unevenly distributed by oscillation phases.

\section{Analytical model}\label{sec:analyt}

\subsection{Beam density and wakefield potential}
\label{s4A}

Our analytical model describing the equilibrium shape of bunch bodies relies on two assumptions.
First, we assume that the period of radial oscillations of the bunch particles is much shorter that the characteristic evolution time of the potential well in which these particles oscillate. 
This is justified if the wakefield amplitude increases along the bunch train and the period of particle oscillations decreases accordingly.
Later bunches oscillate in deeper potential wells with shorter periods, while the evolution time of the potential wells themselves is determined by earlier bunches that evolve over longer periods.
Second, we assume that the radial profiles of bunch density ($n_\perp$) and wakefield potential ($R$) are the same in the beam cross-sections that effectively drive the wave. 
This follows from the previous section.

When developing the analytical model, we use an approach similar to that of Ref.~\onlinecite{PoP24-023119}.

The beam is in radial equilibrium with the wakefield if the radial profiles of beam density $n_\perp(r)$ and wakefield potential $R(r)$ satisfy not only Eq.~\eqref{e7}, but also the relation\cite{PoP24-023119}
\begin{equation}\label{e17}
    n_\perp (r) = \frac{1}{2 \pi k_p r} \int_r^\infty \frac{D(r_a) \, dr_a}{\tilde \tau(r_a) \sqrt{R(r_a) - R(r)}},
\end{equation}
which depends on the distribution $D(r_a)$ of beam particles by oscillation amplitudes $r_a$ and the quantity
\begin{equation}
    \tilde\tau(r_a) = \int_0^{r_a} \frac{dr}{\sqrt{R(r_a)-R(r)}},
\end{equation}
which is proportional to the period of particle oscillations.
Here we assume
\begin{equation}\label{e19}
    \int_0^\infty D(r_a) \, dr_a = 1.
\end{equation}
To obtain Eq.~\eqref{e17}, we note that if the potential well and the particle energy change slowly compared to the period of particle radial oscillations, then the energy of transverse motion is approximately constant during this period:
\begin{equation}\label{e20}
    \frac{\gamma_b m v_r^2}{2} + e \Phi (r_p) =  e \Phi (r_a) = \text{const},
\end{equation}
where $v_r$ is the radial velocity of the particle. 
The oscillation period is
\begin{equation}
    \tau =  4 \int_0^{r_a} \frac{dr}{v_r} = 4 \int_0^{r_a} \frac{\sqrt{\gamma_b m} \, dr}{\sqrt{2e [\Phi(r_a)-\Phi(r)]}},
\end{equation}
and the time $dt$ that the particle spends in a radial interval $dr$ during this period is
\begin{equation}
    dt = \frac{4 \sqrt{\gamma_b m} \, dr}{\sqrt{2e [\Phi(r_a)-\Phi(r)]}}.
\end{equation}
The corresponding ``fraction'' of the particle is 
\begin{equation}
    \frac{dt}{\tau} = \frac{dr}{\tilde\tau(r_a)\sqrt{R(r_a)-R(r)}}, 
\end{equation}
from which formula \eqref{e17} follows.

The relationship~\eqref{e17} depends on the specific type of function $D(r_a)$. 
In Ref.~\onlinecite{PoP24-023119}, this function was taken from simulations and analytically approximated. 
Here, we derive it from the conservation of the adiabatic invariant
\begin{multline}
    J(r_a) = \oint v_r(r') dr' = 4 \int_0^{r_a} \sqrt{\frac{2e [\Phi(r_a)-\Phi(r')]}{\gamma_b m} }\, dr' \\ \label{e24}
    \propto \int_0^{r_a} \sqrt{R(r_a)-R(r')}\, dr'.
\end{multline}
We can find the function $J(r_a)$ if we know the shape of the potential well $R(r)$.

Initially (before the beam evolution begins), all particles are at their turning points, because the initial angular spread of the beam is negligibly small compared to the angular spread gained during equilibration. 
Therefore, for the beam of radial profile \eqref{e1}, initially
\begin{equation}
    D(r_a) = D_0(r_a) \equiv \frac{r_a}{\sigma_r^2} e^{-r_a^2/(2 \sigma_r^2)}.
\end{equation}
By calculating the wakefield potential \eqref{e7} for the initial beam density \eqref{e9} and substituting the resulting function $R(r)$ into Eq.~\eqref{e24}, we find the initial relation between oscillation amplitudes and adiabatic invariants. 
We denote it by $J_0 (r_{a0})$; now it is a known function, and it can be inverted. 

Consider a particle that initially had a turning point at $r_{a0}$.
After changing the potential well, its turning point is at $r_a$. 
Since the adiabatic invariant is conserved, 
\begin{equation}\label{e27}
    J(r_a) = J_0 (r_{a0}),
\end{equation}
and we can relate the initial and final turning points:
\begin{equation}\label{e28}
    r_{a0} = J_0^{-1} (J(r_a)),
\end{equation}
where $J_0^{-1}$ is the inverse function of $J_0 (r_{a0})$.
The  distribution $D(r_a)$ of beam particles by oscillation amplitudes is proportional to the ratio of the number $dN$ of beam particles having turning points in the interval $dr_a$ to the width of this interval. 
We can transform this ratio as follows:
\begin{equation}\label{e29}
    D(r_a) = \alpha \frac{dN}{dr_a} = \alpha \frac{dN}{dr_{a0}} \frac{dr_{a0}}{dr_a} = D_0 (r_{a0}(r_a)) \frac{dr_{a0}}{dr_a}, 
\end{equation}
where $\alpha$ is the normalization coefficient for satisfying condition \eqref{e19}.

Using the above formulas, we can organize an iterative process. 
We start with the initial beam profile \eqref{e9} and functions $R(r)$, $J_0 (r_{a0})$, and $D_0(r_{a0})$ calculated for it.
Then we sequentially find the next iterations: for $n_\perp(r)$ from Eq.~\eqref{e17}, for $R(r)$ from Eq.~\eqref{e7}, for $J(r_a)$ from Eq.~\eqref{e24}, and for $D(r_a)$ from Eqs.~\eqref{e28}--\eqref{e29}. 
The process repeats until the relative difference between successive iterations for $n_\perp(r)$ becomes smaller than a desired value. 
It usually takes 5--7 iterations for the difference to become less than $10^{-3}$.

\begin{figure}[tb]
\includegraphics{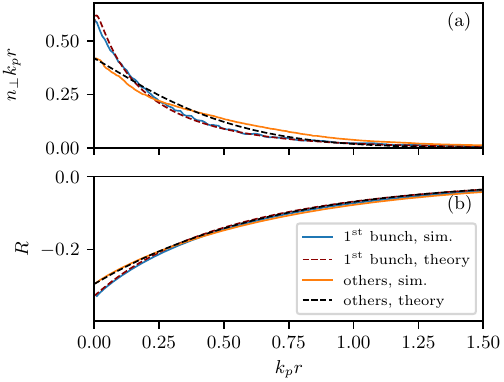}
\caption{ (a) $r$-weighted radial density profiles $n_\perp (r)$ and (b) the corresponding radial profiles $R(r)$ of the wakefield potential obtained analytically (dashed lines) or from numerical simulations by averaging over the bodies of the first or all other bunches.}
\label{fig6-rdrop}
\end{figure}

The theoretical beam density profile $n_\perp(r)$, obtained from the assumption of adiabaticity \eqref{e27}, agrees well with the simulation results if the latter are averaged according to Eq.~\eqref{e10} over all bunch bodies except the first one [Fig.\,\ref{fig6-rdrop}(a)]. 
The corresponding profiles of wakefield potential $R(r)$ also agree [Fig.\,\ref{fig6-rdrop}(b)].

\begin{figure}[tb]
\includegraphics{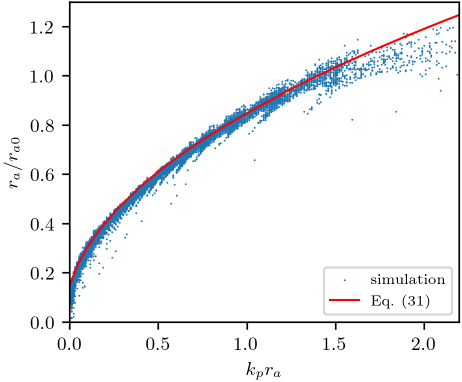}
\caption{ The dependence of the ratio $r_a / r_{a0}$ on the final oscillation amplitude $r_a$ of the particles in the first bunch, obtained from simulations (points) and approximated by  formula \eqref{e31a} (line).}
\label{fig7-sqrt}
\end{figure}

As to the first bunch, its density is more strongly peaked, and the corresponding wakefield potential is deeper on the axis (Fig.\,\ref{fig6-rdrop}). 
This is because the transition of the first bunch to equilibrium is not adiabatic.\cite{PoP24-023119}
The approach described in Ref.~\onlinecite{PoP24-023119}, in which the ratio between the initial and final turning points is approximated using an empirical formula, is more consistent with the simulation results.
An analysis of bunches with various initial radii and longitudinal profiles suggests the formula
\begin{equation}\label{e31a}
    \frac{r_a}{r_{a0}} \approx \sqrt{\frac{r_a}{2.75 \sigma_r}+61 \left( \frac{n_{b0}}{n_0}\right)^{4/3}(k_p \sigma_r)^{2/3}}
\end{equation}
(Fig.\,\ref{fig7-sqrt}), which is more versatile than the one used in Ref.~\onlinecite{PoP24-023119}. 
The second term in Eq.~\eqref{e31a} is added to facilitate comparison with numerical simulations if the beam density $n_{b0}$ is not infinitesimally small. 
It modifies the formula at small radii, where the assumption of a linear plasma response is violated due to the sharp peaking of the beam density. 
The theoretical radial dependencies obtained iteratively using Eq.~\eqref{e31a} instead of Eq.~\eqref{e27} are close to the simulation results for the body of the first bunch (Fig.\,\ref{fig6-rdrop}).

Using Eq.~\eqref{e13}, we can calculate the effective beam current $I_\text{eff}$ assuming that $n_\parallel (\xi)$ remains unchanged and determined by Eq.~\eqref{e9}. 
The peaks of the simulated effective current gradually approach the value calculated for beam bodies, and the peak current of the first bunch is close to the theoretical value for the first bunch [Fig.\,\ref{fig3-general}(b)]. 
Both theoretical lines are almost twice as high as the line calculated for the initial Gaussian distribution, indicating that beam layers become twice as efficient when equilibrium is reached.
For bunches in the front part of the beam, the agreement between theoretical and simulated effective currents is not as good as for other compared quantities. 
This indicates that the motion of near-axis particles, which contribute most  to the effective current due to the singularity of the Bessel function $K_0$, deviates from the adiabatic law more strongly than the motion of other particles.
In turn, this is a consequence of the density singularity on the axis, which violates the linearity of the plasma response.

\begin{figure}[tb]
\includegraphics{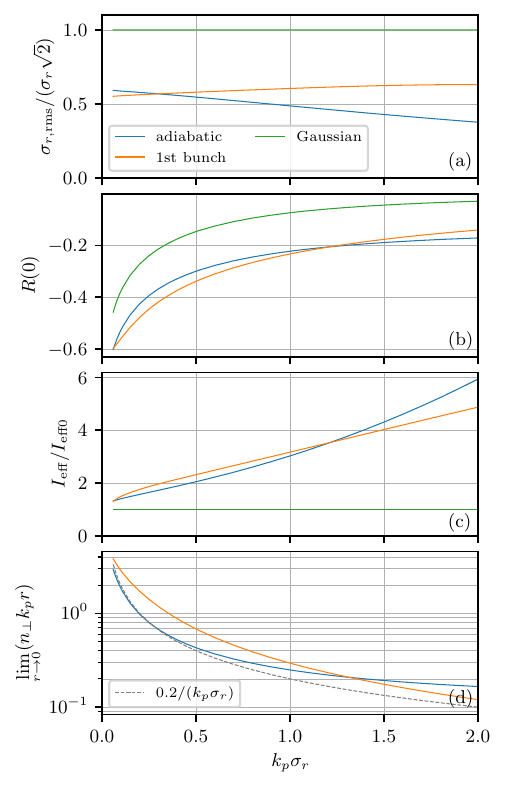}
\caption{ Beam properties calculated on the basis of Eq.~\eqref{e27} (``adiabatic''),  Eq.~\eqref{e31a} (``1st bunch''), and Eq.~\eqref{e9} (``Gaussian''): 
(a) root mean square bunch radius $\sigma_{r, \mathrm{rms}}$ compared to the initial beam radius $\sigma_r$, (b) potential well depth $R(0)$ and (c) its comparison with that for the initial Gaussian beam, (d) degree of beam density peaking. 
The dashed gray line in (d) helps the eye to see the scaling $\propto \sigma_r^{-1}$.}
\label{fig8-parametric}
\end{figure}

With the analytical model, we can easily examine how the parameters of  equilibrium bunches depend on the initial beam radius $\sigma_r$ (Fig.\,\ref{fig8-parametric}). 
As we see, the root mean square radius of the beam decreases by approximately a factor of two in a wide range of $\sigma_r$ [Fig.\,\ref{fig8-parametric}(a)]. 
The bunch contribution to driving the wakefield increases accordingly [Fig.\,\ref{fig8-parametric}(b)], and this gain, compared to the initial Gaussian beam, is greater for wide beams [Fig.\,\ref{fig8-parametric}(c)]. 
This same ratio shows how the effective current increases as a result of beam equilibration.
The degree of density peaking increases proportionally to $\sigma_r^{-1}$ [Fig.\,\ref{fig8-parametric}(d)], so the density of initially narrow beams can easily exceed the plasma density, even if the beam current is low.

\subsection{Phase space density}\label{sec:phase}

Knowing the radial profile of particle density $n_\perp (r)$, we can find the particle density in phase space $\rho(r, p_r, \xi)$. 
This function depends on $\xi$ (Fig.\,\ref{fig5-simphase}, bodies) through the dependence of the potential well depth $\Phi$ and beam density $n_b$ on $\xi$.
To obtain its universal shape, we introduce the normalized profile of the wakefield potential
\begin{equation}\label{e33a}
    \phi(r) = \frac{R(r)}{R(0)}
\end{equation}
and the normalized radial momentum
\begin{equation}\label{e33}
    \tilde{p}_r = \frac{p_r}{p_s},
\end{equation}
where
\begin{equation}\label{e34}
     p_s = |S(0, \xi)| =  m c \sqrt{2 \gamma_b |\Phi(0, \xi)| / \Phi_0}
\end{equation}
is the half-width of the separatrix calculated according to Eqs.~\eqref{e14} and  \eqref{e5} for the analytically obtained function $R(r)$ and an arbitrary function $Z(\xi)$.
Since we assume that the radial dependence of $\Phi(r,\xi)$ is the same in all cross-sections, the function $\Phi(r,\xi)$ is separable. 
Since $R(r)$ is monotonically increasing [Fig.\,\ref{fig6-rdrop}(b)], $\Phi_m (r, \xi) = 0$, and $\phi(r)$ is a positive function that decreases monotonically from 1 to 0.

\begin{figure}[tb]
\includegraphics{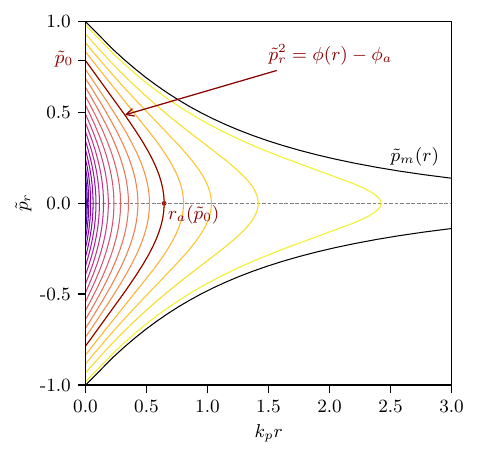}
\caption{ Parametrization of phase trajectories. The trajectories are plotted according to the adiabatic analytical model described in Sec.~\ref{s4A}.}
\label{fig9-phspace}
\end{figure}

The law \eqref{e20} of energy conservation for the normalized quantities becomes
\begin{equation}\label{e32}
    \tilde{p}_0^2 - 1 = \tilde{p}_r^2 - \phi(r),
\end{equation}
where $\tilde p_0$ is the normalized transverse momentum of the particle on the axis.
This equation describes particle trajectories in the phase space $(r, \tilde p_r)$ and also the separatrix
\begin{equation}
    \tilde p_m (r) = \pm \sqrt{\phi(r)}
\end{equation}
(Fig.\,\ref{fig9-phspace}).
The separatrix is the same for all $\xi$, as is the normalized particle density in phase space, which we define as 
\begin{equation}
    \tilde \rho (r, \tilde p_r) = \frac{k_p p_s \rho (r, p_s \tilde p_r, \xi)}{n_{b0} n_\parallel (\xi)}.
\end{equation}
The analogue of Eq.~\eqref{e16} for this function takes the form
\begin{equation}\label{e36}
    \int_{-|\tilde p_m|}^{|\tilde p_m|} \tilde{\rho}(r, \tilde{p}_r)\,d\tilde{p}_r = 2\pi k_p r n_{\perp}(r),
\end{equation}
where we assume that there are no beam particles outside the separatrix.

A phase trajectory \eqref{e32} crosses the horizontal axis ($\tilde p_r = 0$) at the turning point
\begin{equation}
    r_a(\tilde{p}_0) = \phi^{-1}(1-\tilde{p}_0^2),
\end{equation}
where $\phi^{-1}$ is the inverse function of $\phi(r)$.
We denote the wakefield potential at this point as $\phi_a = \phi(r_a)$.
If the phase mixing of the beam is complete, the phase density $\tilde{\rho}$ is constant along this trajectory.
Thus, the phase density is a function of only one parameter that distinguishes phase trajectories. 
This could be $\tilde p_0$, $r_a$, or $\phi_a$.
For further calculations, it is more convenient to distinguish the trajectories by the parameter $\phi_a$:
\begin{equation}
    \tilde{\rho}(r, \tilde{p}_r) = \rho_a(\phi_a (r, \tilde{p}_r)).
\end{equation}
The trajectory \eqref{e32} itself is determined by the equation
\begin{equation}\label{e41}
    \tilde p_r^2 = \phi(r) - \phi_a,
\end{equation}
so for a fixed $r$ we have
\begin{equation}
    \frac{d \tilde p_r}{d \phi_a} = - \frac{1}{2 \sqrt{\phi(r) - \phi_a}}.
\end{equation}
Equation~\eqref{e36} can then be rewritten as
\begin{equation} \label{e40}
    2 \pi k_p r n_\perp (r) = \int_0^{\phi (r)} \frac{\rho_a (\phi_a) \, d \phi_a}{\sqrt{\phi (r) - \phi_a}}.
\end{equation}
This is an integral Abel equation for an unknown function $\rho_a (\phi_a)$.\cite{Abel} Its solution is
\begin{equation}\label{e44}
    \rho_a (\phi_a) = 2 k_p \frac{d}{d \phi_a} \int_0^{\phi_a} \frac{r n_\perp (r) \, d\phi'}{\sqrt{\phi_a - \phi'}} ,
\end{equation}
where $r = \phi^{-1} (\phi')$.

\begin{figure}[tb]
\includegraphics{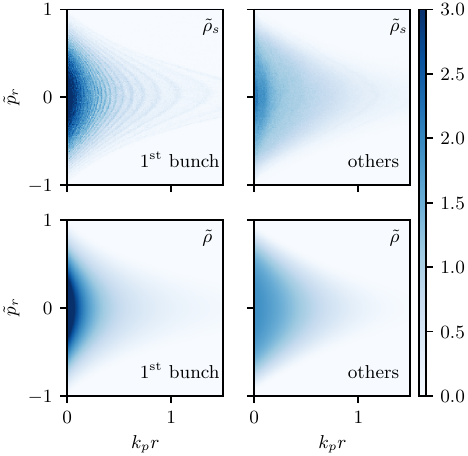}
\caption{ Simulated ($\tilde \rho_s$) and theoretical ($\tilde \rho$) normalized particle densities for bodies of the first and all other bunches.}
\label{fig10-phase}
\end{figure}

\begin{figure}[tb]
\includegraphics{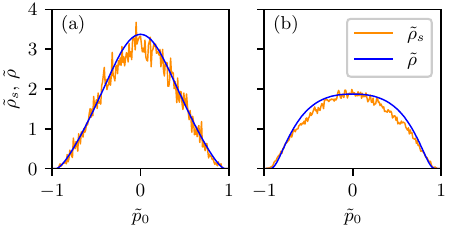}
\caption{ Simulated ($\tilde \rho_s$) and theoretical ($\tilde \rho$) phase densities on phase trajectories with different $\tilde p_0$ for bodies of the first (a) and all other (b) bunches. }
\label{fig11-pr}
\end{figure}

In simulations, the beam density $n_b(r, \xi)$ is not separable, and the particle density in phase space fluctuates depending on $\xi$ because of incomplete phase mixing and a limited number of macroparticles in the beam (Fig.\,\ref{fig5-simphase}).
To compare the simulation results with theory, we must represent the phase density as a function of normalized momentum $\tilde p_r$, with the potential $\Phi(0,\xi)$ in formula \eqref{e34} taken from the simulation, and average it over a wide interval of $\xi$: 
\begin{equation}\label{e45}
    \tilde \rho_s (r, \tilde{p}_r) = A \int \rho (r, p_s \tilde{p}_r, \xi') \,d\xi',
\end{equation}
where the coefficient $A$ is the same as in Eq.~(\ref{e10}).
With the momentum normalization, the density distributions patterns become quite similar in all cross-sections of bunch bodies.
When averaged, the simulated patterns are very close to the theoretical ones for bodies of both the first bunch and other bunches (Fig.\,\ref{fig10-phase}).
Cross-sections of the phase density maps at $r=0$ show us the phase densities on individual phase trajectories (Fig.\,\ref{fig11-pr}).
They also agree reasonably.

\section{Approximate expressions}
\label{sec:approx}

The radial profiles of $n_\perp (r)$ and $R(r)$ (Fig.\,\ref{fig6-rdrop}) that we obtained iteratively are quite accurate. 
However, we cannot write explicit expressions for them, which makes them inconvenient to use. 
In this Section, we approximate the density profiles with elementary functions, which allows us to derive relatively simple formulas for other characteristics of the equilibrium beams.

\begin{figure}[tb]
\includegraphics{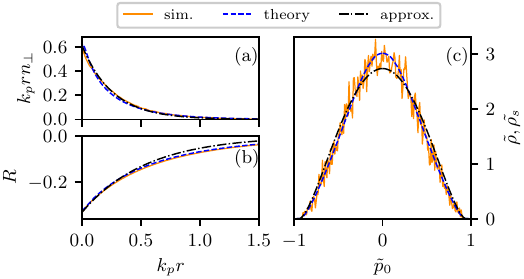}
\caption{ Comparison of the approximate expressions with simulation results and iterative theoretical model for the first bunch.}
\label{fig12-engf}
\end{figure}

\begin{figure}[tb]
\includegraphics{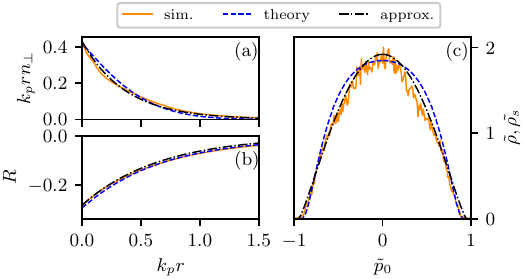}
\caption{ Comparison of the approximate expressions with simulation results and iterative theoretical model for bodies of all bunches except the first one.}
\label{fig13-engb}
\end{figure}

\begin{figure}[tb]
\includegraphics{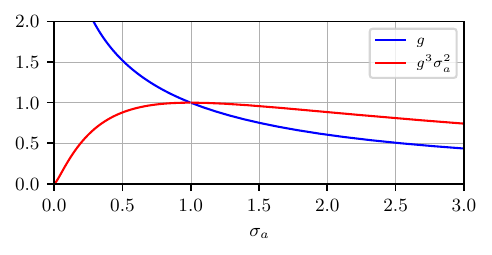}
\caption{Functions used in approximate expressions for the plasma response.}
\label{fig14-g}
\end{figure}

The function
\begin{equation}\label{e46}
    n_\perp (r) = \frac{1}{2 \pi k_p r \sigma_a} e^{-k_p r/\sigma_a},
\end{equation}
with 
\begin{equation}
    \sigma_a = k_p \sigma_r / 2
\end{equation}
for the first bunch and
\begin{equation}
    \sigma_a = k_p \sigma_r / \sqrt{2}
\end{equation}
for bodies of other bunches provides a fairly accurate approximation of the density profiles [Figs.\,\ref{fig12-engf}(a) and \ref{fig13-engb}(a)].
Substituting this to Eq.~\eqref{e7} and assuming $k_p r \ll 1$, $k_p r \ll \sigma_a$, after some math detailed in Appendix~\ref{app1}, we obtain
\begin{equation}\label{e49}
    R(r) \approx -\frac{g}{2\pi} \exp \left( -\frac{k_p r}{\sigma_a g}\right),
\end{equation}
where
\begin{equation}\label{e51}
    g \equiv g(\sigma_a) = \left\{ \begin{array}{ll}
    \displaystyle \frac{\operatorname{acosh}{(1/\sigma_a)}}{\sqrt{1-\sigma_a^2}}, & 0 \leq \sigma_a \leq 1, \\
    \displaystyle \frac{\operatorname{arccos}{(1/\sigma_a)}}{\sqrt{\sigma_a^2-1}}, & \sigma_a > 1
    \end{array}\right.
\end{equation}
(Fig.\,\ref{fig14-g}).
The expression \eqref{e49} does not strictly satisfy the normalization condition \eqref{e8a}: The left-hand-side of Eq.~\eqref{e8a} equals $\sigma_a^2 g^3$ in absolute value. 
However, it is close to unity in a wide range of $\sigma_a$ (Fig.\,\ref{fig14-g}). Therefore, expression~\eqref{e49} provides good agreement over a fairly wide range of $r$ [Figs.\,\ref{fig12-engf}(b) and \ref{fig13-engb}(b)], even though it was derived for small radii.

According to Eq.~\eqref{e41}, the phase trajectories are given by the formula
\begin{equation}
    \tilde p_r^2 \approx e^{-k_p r/(\sigma_ag)} - e^{-k_p r_a/(\sigma_ag)}.
\end{equation}
At small $r$, they have a parabolic shape (Fig.\,\ref{fig9-phspace}):
\begin{equation}
    \tilde p_r^2 \approx \frac{r_a - r}{\sigma_ag}.
\end{equation}

Substituting formulas \eqref{e46} and \eqref{e49} into Eq.~\eqref{e44}, we obtain the phase density (Appendix~\ref{app2}):
\begin{equation}\label{e54a}
    \rho_a(\phi_a) = \frac{\Gamma(g+1)}{\sqrt{\pi} \sigma_a \Gamma(g+1/2)} \phi_a^{g-1/2},
\end{equation}
where $\Gamma$ is the gamma function. 
The agreement is also good [Figs.\,\ref{fig12-engf}(c) and \ref{fig13-engb}(c)].

\section{Discussion}
\label{sec:discuss}

The derived formulas are not universal, and this is most clearly evident when comparing the theory with the simulated effective current in Fig.\,\ref{fig3-general}(b).
There are areas where the adiabatic model yields inaccurate results (the first few bunches) or does not work at all (the tails and heads of the bunches).
The inaccuracies arise due to different ratios between the time scales of betatron oscillations of individual particles and the evolution time of the field as a whole.
In the first bunch, the evolution of the field is governed by the particles that oscillate in it; therefore, the time scales are the same,\cite{PoP24-023119} and the empirical model is valid.
While the first $\sim$10 bunches are reaching equilibrium, the adiabaticity condition holds only approximately (the timescales are comparable), so their effective current is smaller.
This shows that non-adiabaticity of bunch compression can not only increase the density peaking but also decrease it.
In turn, in the second half of the simulated bunch train, a situation arises in which the focusing field acting on the bunches is strong, and the particles oscillate rapidly in it. The evolution time of this field is determined by the preceding bunches, which are located in weaker fields and evolve more slowly, so our adiabatic model is accurate.

Curiously, there is another deviation from the developed model at the rear of the bunch train.
If we examine the evolution of individual bunches (as shown in Fig.\,10 of Ref.~\onlinecite{PoP22-103110}), we can see that during their formation, these bunches are in a weak focusing field (for this particular plasma density profile), and the adiabaticity condition holds only approximately for them.
For the same reason, the bunch tails are not described analytically.
As they evolve, they go through a period of defocusing, so the assumption of adiabaticity does not apply to them, even approximately.

Overall, however, the model provides a fairly accurate description of the bunch parts that drive the wave.
Consequently, the better we shape the bunches by controlling the plasma density profile, the more accurately our model will describe them.

\section{Summary}
\label{sec:summ}

Upon reaching transverse equilibrium with a linear wakefield, the self-modulated beam changes its radial density profile from the initial Gaussian distribution to a more peaked one, characterized by a density singularity on the axis. 
The radial distribution of the wakefield potential acquires a corresponding funnel-like shape. 
In this equilibrium state, the beam excites the wakefield several times more efficiently than in the case of the original Gaussian shape.

The longitudinal structure of the bunches is not uniform. 
There is a difference in shape between the leading parts of the bunches (``bodies''), which drive the wave, and the trailing parts (``tails''), which are accelerated by the wave.
The equilibrium state of the bunch bodies is well described by an analytical model based on the conservation of the adiabatic invariant. 
The model predicts the radial profiles of the beam density and wakefield potential, as well as the particle distribution in the transverse phase space, and shows good agreement with numerical simulations.

The fractions of the beam charge that are decelerated or accelerated determine the efficiency of driving the wave.
They depend on the longitudinal density profile of the plasma at the self-modulation stage.
For efficient drivers created with a stepped-up plasma density profile, most of the beam charge is concentrated in the ``bodies'', so most of the beam is analytically tractable.
However, plasma density profiles that would allow for the complete elimination of the accelerated bunch parts have not yet been found, if this is even possible.

While the adiabatic model accounts for most of the bunch train, the first bunch requires a different approach due to the non-adiabatic nature of its formation. 
The model describing the first bunch is based on a relationship empirically obtained from simulations, but it also accurately describes the equilibrium state of the bunch.

The developed models rely on iterative processes and numerical integrations, which can make them difficult to apply. 
There are also approximate formulas based on elementary functions, which provide a practical method for calculating radial profiles and phase densities. 
These expressions are valid over a wide range of initial beam radii and eliminate the need for iterative numerical procedures.

\acknowledgements

The analytical part of this work was supported by the Russian Science Foundation, project No. 23-12-00028.
The simulation part of the work (Sec.~\ref{sec:phenom}) was supported by Budker INP with ongoing institutional funding. Numerical simulations were performed on HPC-cluster ``Akademik V.\,M.\,Matrosov''\cite{Matrosov}.

\appendix

\section{Normalization of $R(r)$}
\label{app0}

Let us substitute Eq.~\eqref{e7} into the left-hand side of Eq.~\eqref{e8a}:
\begin{multline}
    - 2 \pi k_p^4 \int_0^\infty r \, dr \left( \int_0^r r' \, dr' I_0 (k_p r') K_0 (k_p r) \, n_\perp (r') \right. \\ 
    \left.+ \int_r^\infty r' \, dr' I_0 (k_p r) K_0 (k_p r') \, n_\perp (r') \right),
\end{multline}
and change integration order:
\begin{multline}
    - 2 \pi k_p^4 \biggl( \int_0^\infty r' \, dr' \int_{r'}^\infty r \, dr \,I_0 (k_p r') K_0 (k_p r) \, n_\perp (r') \\ 
    + \int_0^\infty r' \, dr' \int_0^{r'} r \, dr \, I_0 (k_p r) K_0 (k_p r') \, n_\perp (r') \biggr).
\end{multline}
Since
\begin{gather}
    k_p^2 \int_{r'}^\infty r \, dr \, K_0 (k_p r) = k_p r' K_1(k_p r'), \\
    k_p^2 \int_0^{r'} r \, dr I_0 \, (k_p r) = k_p r' I_1(k_p r'),
\end{gather}
and
\begin{equation}
    I_0 (k_p r') K_1 (k_p r') + I_1 (k_p r') K_0 (k_p r') = \frac{1}{k_p r'},
\end{equation}
we obtain Eq.~\eqref{e8a}:
\begin{multline}
    - 2 \pi k_p^4 \int_0^\infty r' \, dr' \, n_\perp (r') \biggl( I_0 (k_p r') \int_{r'}^\infty r \, dr \, K_0 (k_p r) \\ 
    + K_0 (k_p r') \int_0^{r'} r \, dr \, I_0 (k_p r) \biggr) \\
    =  - 2 \pi k_p^2 \int_0^\infty r' \, dr' \, n_\perp (r') = -1.
\end{multline}

\section{Approximation for $R(r)$}
\label{app1}

The function $R(r)$ resembles an exponential function (Fig.\,\ref{fig6-rdrop}), so we look for it in this particular form:
\begin{equation}\label{e55}
    R(r) \approx a e^{b r}.
\end{equation}
We choose the coefficients $a$ and $b$ such that to obtain the correct values of $R$ and its derivative at $r=0$:
\begin{equation}\label{e57}
    a = R(0), \qquad  b = \frac{1}{R(0)} \left. \frac{d R}{dr}\right|_{r=0}.
\end{equation}
Substituting the radial density distribution \eqref{e46} into Eq.~\eqref{e7},
\begin{gather}
    \nonumber
    R(r) = - \frac{k_p}{2 \pi \sigma_a} \int_0^r d r' I_0 (k_p r') K_0 (k_p r) e^{-k_p r'/\sigma_a}\,  \\ \label{e54}
        - \frac{k_p}{2 \pi \sigma_a} \int_r^\infty d r' I_0 (k_p r) K_0 (k_p r') \, e^{-k_p r'/\sigma_a},
\end{gather}
we find
\begin{equation}
    R(0) = - \frac{k_p}{2 \pi \sigma_a} I_0 (0) \int_0^\infty d r'  K_0 (k_p r') \, e^{-k_p r'/\sigma_a}.
\end{equation}
This integral can be expressed\cite{GR} as
\begin{equation}\label{e59}
    R(0) = -\frac{g(\sigma_a)}{2\pi}
\end{equation}
with $g(\sigma_a)$ given by Eq.~\eqref{e51}.

To find $dR/dr$, we assume $r \ll \min(k_p^{-1}, \sigma_a/k_p)$ and use series representations of the modified Bessel functions. 
To within terms linear in $r$, we have
\begin{multline} \label{e54}
    R(r) = R(0) - \frac{k_p K_0 (k_p r)}{2 \pi \sigma_a} \int_0^r d r' I_0 (k_p r')  e^{-k_p r'/\sigma_a}\,  \\ 
        + \frac{k_p I_0 (k_p r)}{2 \pi \sigma_a} \int_0^r d r' K_0 (k_p r') \, e^{-k_p r'/\sigma_a} \\
    \approx R(0) - \frac{k_p (-\ln(k_p r/2) - C)}{2 \pi \sigma_a} \, r \\
        + \frac{k_p}{2 \pi \sigma_a} \int_0^r d r' (-\ln(k_p r'/2) - C) \\
    = R(0) + \frac{k_p r}{2 \pi \sigma_a},
\end{multline}
where $C$ is Euler's constant. Therefore,
\begin{equation}
    a = -\frac{g}{2\pi}, \qquad b = -\frac{k_p}{\sigma_a g},
\end{equation}
from which the expression \eqref{e49} follows.

\section{Approximation for phase density}
\label{app2}

To find an approximate formula for the particle density in phase space, we need to express the integrand of Eq.~\eqref{e44} in terms of the normalized wakefield potential $\phi$ \eqref{e33a} using the available approximations for $n_\perp(r)$ \eqref{e46} and $R(r)$ \eqref{e49}:
\begin{gather}
    \phi = \exp \left( -\frac{k_pr}{\sigma_a g} \right),  \qquad
    r = -\frac{\sigma_a g}{k_p} \ln{\phi},\\
    n_\perp(\phi) = -\frac{1}{2\pi \sigma_a^2 g \ln{\phi}}\phi^g.
\end{gather}
After substitution, Eq.~\eqref{e44} becomes
\begin{equation}
    \rho_a(\phi_a) = \frac{1}{\pi \sigma_a} \frac{d}{d\phi_a} \int_0^{\phi_a} \frac{\phi^g d\phi}{\sqrt{\phi_a - \phi}}.
\end{equation}
By introducing a new variable $y = \phi/\phi_a$,
\begin{equation}\label{e65a}
    \rho_a(\phi_a) = \frac{1}{\pi \sigma_a}\frac{d \phi_a^{g+1/2}}{d\phi_a}  \int_0^{1} y^g (1-y)^{-1/2}dy.
\end{equation}
we reduce the integral to the beta function $\mathbf B$ (Euler's integral of the first kind):
\begin{equation}
    \int_0^{1} \frac{y^g\,dy}{\sqrt{1-y}} = \mathbf{B}(g+1, 1/2) = \frac{\Gamma (g+1) \Gamma(1/2)}{\Gamma (g+3/2)}.
\end{equation}
After differentiating in Eq.~\eqref{e65a} and simplifying the gamma functions, we obtain Eq.~\eqref{e54a}:
\begin{equation}
    \rho_a(\phi_a) = \frac{\Gamma(g+1)}{\sqrt{\pi} \sigma_a \Gamma(g+1/2)} \phi_a^{g-1/2}.
\end{equation}

\end{document}